\documentclass[conference]{IEEEtran}
\IEEEoverridecommandlockouts
\usepackage{cite}
\usepackage{amsmath,amssymb,amsfonts}
\usepackage{algorithmic}
\usepackage{graphicx}
\usepackage{textcomp}
\usepackage{xcolor}
\def\BibTeX{{\rm B\kern-.05em{\sc i\kern-.025em b}\kern-.08em
    T\kern-.1667em\lower.7ex\hbox{E}\kern-.125emX}}

\usepackage{comment}
\usepackage[hidelinks]{hyperref}

\newcommand{\bbC}{\mathbb{C}}
\newcommand{\rmd}{\mathrm{d}}
\newcommand{\bbE}{\mathbb{E}}\newcommand{\rme}{\mathrm{e}}

\newcommand{\bbH}{\mathbb{H}}

\newcommand{\bbR}{\mathbb{R}}

\newcommand{\bfA}{\mathbf{A}}\newcommand{\bfa}{\mathbf{a}}

\newcommand{\sfC}{\mathsf{C}}

\newcommand{\bfH}{\mathbf{H}}\newcommand{\sfH}{\mathsf{H}}
\newcommand{\sfI}{\mathsf{I}}
\newcommand{\bfJ}{\mathbf{J}}

\newcommand{\bfR}{\mathbf{R}}

\newcommand{\bfT}{\mathbf{T}}\newcommand{\sfT}{\mathsf{T}}

\newcommand{\bfX}{\mathbf{X}}\newcommand{\bfx}{\mathbf{x}}\newcommand{\sfX}{\mathsf{X}}
\newcommand{\bfY}{\mathbf{Y}}\newcommand{\bfy}{\mathbf{y}}\newcommand{\sfY}{\mathsf{Y}}
\newcommand{\bfZ}{\mathbf{Z}}\newcommand{\sfZ}{\mathsf{Z}}

\newcommand{\cA}{\mathcal{A}}

\newcommand{\cX}{\mathcal{X}}
\newcommand{\sfy}{\mathsf{y}}\newcommand{\cY}{\mathcal{Y}}

\newcommand{\expect}[1]{{\mathbb{E}}\left[#1\right]}
\newcommand{\expcnd}[2]{{\mathbb{E}}\left[ #1 \;\middle|\; #2\right]}

\newcommand{\variance}[1]{{\mathsf{Var}}\left[#1\right]}
\newcommand{\varcnd}[2]{{\mathsf{Var}}\left[ #1 \;\middle|\; #2\right]}

\newcommand{\covariance}[1]{{\mathsf{Cov}}\left[#1\right]}

\newcommand{\vect}[1]{{\text{vec}}\left(#1\right)}

\newcommand{\tr}[1]
{{\text{tr}}\left(#1\right)}

\newtheorem{lemma}{Lemma}

\newtheorem{prop}{Proposition}

\newtheorem{theorem}{Theorem}

\newcommand{\dt}[1]{{\red #1}}

\begin{document}

\title{
MMSE Channel Estimation in Fading MIMO Gaussian Channels With Blockage: A Novel Lower Bound via Poincaré Inequality
}


\author{
\IEEEauthorblockN{Mohammadreza Bakhshizadeh Mohajer$^{*}$, Luca Barletta$^{*}$,  Daniela Tuninetti$^{\dagger}$, Alessandro Tomasoni$^{**}$, \\ Daniele Lo Iacono$^{**}$, and Fabio Osnato$^{**}$} 
$^{*}$ Politecnico di Milano, 20133 Milano, Italy. Email: $\{$mohammadreza.bakhshizadeh, luca.barletta$\}$@polimi.it \\
$^{\dagger}$ University of Illinois Chicago, 60607 Chicago, IL, USA. Email: danielat@uic.edu \\
$^{**}$ STMicroelectronics Srl, 20864 Agrate B.za, Italy. Email: \texttt{firstname}.\texttt{lastname}@st.com}

\maketitle

\begin{abstract}
Integrated sensing and communication is regarded as a key enabler for next-generation wireless networks. 
To optimize the transmitted waveform for both sensing and communication, various performance metrics must be considered. 
This work focuses on sensing, and specifically on the mean square error (MSE) of channel estimation. 
Given the complexity of deriving the MSE, the Bayesian Cramér-Rao Bound (BCRB) is commonly recognized as a lower bound on the minimum MSE. 
However, the BCRB is not applicable to channels with discrete or mixed distributions. 
To address this limitation, a new lower bound based on a Poincaré inequality is proposed and applied to 
fading MIMO AWGN channels with blockage probability, and the behavior of the lower bound at high SNR is precisely characterized.
\end{abstract}


\section{Introduction}
Next-generation wireless networks, such as 
6G, are expected to play a crucial role in supporting various emerging applications. In addition to high-quality connectivity, precise and robust sensing is expected to play a significant role in these networks~\cite{ISAC_survey,5G_6G_ISAC}. Integrated sensing and communication (ISAC), which performs wireless communication and sensing using a single hardware platform and a common waveform, is considered a key technology in 6G with the benefit of reducing hardware cost, size, and power consumption~\cite{6G_survey_ISAC, ISAC_Caire}. 

From the communication point of view, given the channel the objective is to design an optimal transmission signal that maximizes the information rate. However, from the sensing perspective, the problem is to estimate channel parameters knowing the transmitted signal. This leads to different performance metrics and design criteria~\cite{ISAC_Caire}. Future 6G vehicle-to-everything (V2X) is foreseen to explore millimeter wave (mmWave)~\cite{V2X}. Also, the mmWave bands are envisioned to be exploited for sensing in 6G, wireless local area network (WLAN), and next-generation wireless networks~\cite{5G_6G_ISAC}. As a challenge, propagation in these frequencies suffer higher losses and can easily be blocked by objects in the path of the transmitted signal~\cite{UAV_blockage}.

Motivated by ISAC systems and propagation challenges in high frequencies, we focus on characterizing the performance of the sensing subsystem, which is typically done via the estimation mean square error (MSE). Since an exact evaluation of the MSE is challenging, previous work has focused on lower bounding the MSE of weakly unbiased estimators by the Bayesian Cramér-Rao bound (BCRB)~\cite{ISAC_Caire}. 
The BCRB is not tight in general and does not hold for discrete or mixed distributions \cite{Poincare_dytso}. In this work, we propose an alternative lower bound on the MSE of channel estimation based on a Poincaré inequality derived in~\cite{Poincare_dytso}. Specifically, we extend the inequality to cope with the non-canonical exponential family of distributions. Further, we compute the lower bound for Gaussian channels with blockage probability, for both the scalar and vector cases. In both instances, as for the BCRB, our bound is a function of the sample covariance matrix of the transmitted signal, and the exact behavior in the high signal-to-noise ratio (SNR) regime can be characterized. Finally, through numerical evaluations, we compare our bound with the minimum MMSE (MMSE) and the linear MMSE (LMMSE).
\paragraph*{Paper Organization}
In Sec.~\ref{Section:Poincare_lower_bound}, we introduce the Poincaré Lower Bound for the non-canonical exponential family. The calculation of the bound for the scalar channel model with blockage probability and its extension to the vector case is studied in Sec.~\ref{Section:channel_with_blockage}. The same section contains the 
numerical evaluations of the MMSE, its upper bound LMMSE, and the Poincaré lower bound. Finally, Sec.~\ref{Section:conclusion} concludes the paper.

\paragraph*{Notations}
Throughout this paper, deterministic scalar quantities are denoted by lowercase letters, random vectors are denoted by uppercase sans serif letters, deterministic vectors are denoted by bold lowercase letters, and random matrices by bold uppercase letters (e.g. $x, \sfX, \bfx, \bfX$). $\langle \cdot, \cdot\rangle$ denotes the inner product. For a matrix $\bfA$, $\bfA^{\sfT}$, $\bfA^{\dagger}$, $\bfA^{-1}$, $\det(\bfA)$, and $\text{tr}(\bfA)$ denote the transpose, the Hermitian transpose, the inverse, the determinant and the trace of the matrix $\bfA$, respectively. $(\bfA)^{+} = (\bfA^{\sfT} \bfA)^{-1} \bfA^{\sfT}$ denotes the pseudoinverse of matrix $\bfA$. For matrix $\bfA \in \bbR^{n\times m}$, $\text{vec}(\bfA) = [\bfa_{1}^{\sfT}, \bfa_{2}^{\sfT}, \dots , \bfa_{m}^{\sfT}]^{\sfT} \in \bbR^{nm \times 1}$, where $\bfa_i$ is the $i$-th column of $\bfA$, is the vectorization operator. $\sfI_{k}$ is the identity matrix of dimension $k$, $\bf0$ is the column vector of all zeros, and $\delta(\cdot)$ denotes the Kronecker delta. The smallest eigenvalue and the smallest singular value are denoted by $\lambda_{\text{min}}(\bfA)$ and $\sigma_{\min}(\bfA)$, respectively. $\succeq$ denotes the semidefinite ordering and the Kronecker product is denoted by $\otimes$. $\nabla_{\bfx}(\cdot)$, $\bfJ_{\bfx}(\cdot)$, and $\bbH_{\bfx}(\cdot)$ denote the gradient, the Jacobian matrix, and the Hessian matrix with respect to $\bfx$. We define the MMSE of estimating the random variable $\sfX$ given observation $\sfY$ as
\begin{equation}\label{eq:definition_MMSE}
    \text{mmse}(\sfX|\sfY) \triangleq \bbE \left[\|\sfX-\bbE[\sfX|\sfY]\|^{2} \right],
\end{equation}
where $\bbE[\cdot]$ denotes the expectation operator and $\|\cdot \|$ is the Euclidean norm. Finally, the LMMSE is defined as
\begin{align}
    \text{lmmse}(\sfX|\sfY) &\triangleq \variance{\sfX} - \covariance{\sfX,\sfY} \variance{\sfY}^{-1} \covariance{\sfY, \sfX},
\end{align}
where $\variance{\cdot}$ and $\covariance{\cdot}$ are the variance and covariance operators defined as follows
\begin{align}
    \variance{\sfX} = \bbE \left[(\sfX-\bbE[\sfX])(\sfX-\bbE[\sfX])^{\dagger} \right],\\
    \covariance{\sfX,\sfY} = \bbE \left[(\sfX-\bbE[\sfX])(\sfY-\bbE[\sfY])^{\dagger} \right].
\end{align}
\section{Poincaré Lower Bound}
\label{Section:Poincare_lower_bound}
Here, we briefly review the non-canonical exponential family and the Poincaré inequality. Then we extend the alternative MMSE formula and the Poincaré lower bound presented in~\cite{Poincare_dytso} to the more general case of non-canonical exponential family.
\subsection{Non-Canonical Exponential Family}
We consider probability models $\{P_{\sfY | \sfX = \bfx}, \bfx \in \cX \subseteq \bbR^d\}$ supported on $\bfy \in \cY \subseteq \bbR^{k}$ of the non-canonical exponential family with probability density function (pdf)
\begin{align}
    f_{\sfY | \sfX}(\bfy | \bfx) = h(\bfy) \rme^{\langle \eta(\bfx), \bfT(\bfy) \rangle - \phi(\bfx)} \label{eq:exp_family}
\end{align}
where $h: \cY \to [0, \infty)$ is the \emph{base measure}; $\eta: \cX \to \bbR^d$ is the \emph{natural parameter} function; $\bfT: \cY \to \bbR^d$ is the \emph{sufficient statistic} function; $\phi: \cX \to \bbR$ is the \emph{log-partition} function.
Let us also introduce the \emph{information density} function
\begin{align*}
    i_{P_{\sfX,\sfY}}(\bfx, \bfy) &\triangleq \log \frac{\rmd P_{\sfX,\sfY}}{\rmd(P_{\sfX}\cdot P_{\sfY})}(\bfx,\bfy)\nonumber\\
    &= \log \frac{\rmd P_{\sfY|\sfX=\bfx}}{\rmd P_{\sfY}}(\bfy), 
\end{align*}
where $\bfx \in \cX, \, \bfy \in \cY$, and $\frac{\rmd P_{\sfX,\sfY}}{\rmd(P_{\sfX}\cdot P_{\sfY})}(\bfx,\bfy)$ is the Radon-Nikodyme derivative.
Next, we show a result about the gradient of the information density.
\begin{prop}
For the exponential family defined in~\eqref{eq:exp_family} we have that
\begin{align}
    \nabla_{\bfy} i_{P_{\sfX,\sfY}}(\bfx, \bfy) = \bfJ_{\bfy} \bfT(\bfy) \cdot (\eta(\bfx) -\expcnd{\eta(\sfX)}{\sfY=\bfy}),
\end{align}
where $\bfx \in \cX$ and $\bfy \in \cY$.
\label{prop_1}
\end{prop}
\begin{IEEEproof}
Fix some $\bfx \in \cX$. Then,
\begin{align}
    &\nabla_{\bfy} i_{P_{\sfX,\sfY}}(\bfx, \bfy) \nonumber\\
    &= \nabla_{\bfy} \log h(\bfy) + \nabla_{\bfy} \langle \eta(\bfx), \bfT(\bfy)\rangle - \nabla_{\bfy} \log f_{\sfY}(\bfy) \label{eq:use_exp_family} \\
    &= \bfJ_{\bfy} \bfT(\bfy) \cdot \eta(\bfx) -\nabla_{\bfy} \log \frac{f_{\sfY}(\bfy)}{h(\bfy)} \label{eq:gradient_Jacobian_relationship} \\
    &= \bfJ_{\bfy} \bfT(\bfy) \cdot \eta(\bfx) - \bfJ_{\bfy} \bfT(\bfy) \cdot \expcnd{\eta(\sfX)}{\sfY=\bfy} \label{eq:use_TRE} \\
    &= \bfJ_{\bfy} \bfT(\bfy) \cdot (\eta(\bfx) -\expcnd{\eta(\sfX)}{\sfY=\bfy}),
\end{align}
where \eqref{eq:use_exp_family} follows from~\eqref{eq:exp_family}; \eqref{eq:gradient_Jacobian_relationship} follows from the gradient-Jacobian relationship $\nabla_{\bfy} \langle \eta(\bfx), \bfT(\bfy)\rangle=\bfJ_{\bfy} \bfT(\bfy) \cdot \eta(\bfx)$; and in \eqref{eq:use_TRE} we used 
\begin{equation}
    \nabla_{\bfy} \log \frac{f_{\sfY}(\bfy)}{h(\bfy)} = \bfJ_{\bfy} \bfT(\bfy) \cdot \expcnd{\eta(\sfX)}{\sfY=\bfy},
\end{equation}
which is the Tweedie-Robbins-Esposito (TRE) formula~\cite{TRE_identity}.
\end{IEEEproof}
\subsection{Conditional Poincaré Inequality}
In order to present our new lower bound on the MMSE, we first introduce the conditional Poincaré inequality. For a class of functions $\cA$, a conditional probability $P_{\sfY|\sfX=\bfx}$ (for a fixed $\bfx \in \cX$) is said to satisfy a Poincaré inequality with respect to $\cA$ with a constant $\kappa(\bfx) \geq 0$ if for all $f \in \cA$
\begin{align}
    \mathsf{Var}[f(\sfY) | \sfX = \bfx] \leq \frac{1}{\kappa(\bfx)} \expcnd{\|\nabla f(\sfY)\|^2}{\sfX = \bfx}. \label{eq:Poincare_ineq}
\end{align}
There exist several sufficient conditions on $\cA$ and $P_{\sfY|\sfX=\bfx}$ that guarantee the Poincaré inequality holds and identifies the constant $\kappa(\bfx)$. 
Following~\cite{Poincare_dytso}, for a class $\cA$ of continuously differentiable functions, the Bakry-Émery condition defines
\begin{align}
    \kappa_{\text{BE}}(\bfx) = \max{\left\{ \kappa: \bbH_{\bfy} \log \left(\frac{1}{f_{\sfY | \sfX}(\bfy | \bfx)} \right) \succeq \kappa \sfI_{k}, \forall \bfy \in \cY \right\}}.
\end{align}
The Bakry-Émery constant for the non-canonical exponential family is given by the next proposition.
\begin{prop}\label{prop:BE_constant}
    Assume that $P_{\sfY | \sfX}$ has a pdf of the form in~(\ref{eq:exp_family}). Then for $\bfx \in \cX$, we have
    \begin{align}
        \kappa_{\text{BE}}(\bfx) &= \max{\left\{0,\Tilde{\kappa}_{\text{BE}}(\bfx)\right\}}, \\
        \Tilde{\kappa}_{\text{BE}}(\bfx) &= \min_{\bfy \in \cY} \lambda_{\text{min}} \left( \bbH_{\bfy} \log \left( \frac{1}{h(\bfy)} \right) - \bbH_{\bfy} \langle \eta(\bfx) , \bfT (\bfy) \rangle \right).
    \end{align}
\end{prop}
\begin{IEEEproof}
    We have
    \begin{align}
        &\bbH_{\bfy} \log \left( \frac{1}{f_{\sfY | \sfX}(\bfy | \bfx)}\right) \nonumber\\
        &= - \bbH_{\bfy} (\log(h(\bfy)) - \bbH_{\bfy} (\langle \eta(\bfx) , \bfT (\bfy)\rangle ) - \phi(\bfx)) \\
        &\succeq \lambda_{\text{min}} \left( \bbH_{\bfy} \log \left( \frac{1}{h(\bfy)} \right) - \bbH_{\bfy} \langle \eta(\bfx) , \bfT (\bfy) \rangle \right) \sfI_{k},
    \end{align}
    which concludes the proof.
\end{IEEEproof}
\subsection{MMSE and the Poincaré Lower Bound}
In~\cite{Poincare_dytso}, an alternative expression of the estimation MMSE was proposed. Next, we extend that MMSE result for the non-canonical exponential family of distributions.
\begin{theorem}\label{Theorem_mmse_Jacobian}
Assume that $P_{\sfY | \sfX}$ has a pdf of the form in~\eqref{eq:exp_family} and that $\bfJ_{\bfY} \bfT (\sfY)$ has full column rank almost surely (a.s.)~$\sfY$. Then,
\begin{align}
    \text{mmse}(\eta(\sfX) | \sfY) = \expect{\|(\bfJ_{\sfY} \bfT (\sfY))^{+} \nabla_{\sfY} i_{P_{\sfX,\sfY}}(\sfX, \sfY)\|^2}.
    \label{eq:conditionalMMSEdef}
\end{align}
\end{theorem}
\begin{IEEEproof}
    Since $\bfJ_{\sfY} \bfT (\sfY)$ has full rank a.s.~$\sfY$, the pseudo-inverse $(\bfJ_{\sfY} \bfT (\sfY))^{+}$ exists a.s.~$\sfY$. Using Proposition~\ref{prop_1}, we have a.s. $\sfY$,
    \begin{align}
        (\eta(\sfX) - \expcnd{\eta(\sfX)}{\sfY}) = (\bfJ_{\sfY} \bfT (\sfY))^{+} \nabla_{\sfY} i_{P_{\sfX,\sfY}}(\sfX, \sfY).
        \label{eq:new_mmse}
    \end{align}
    By taking the squared norm  and expectation on both sides of~\eqref{eq:new_mmse}, and recalling~\eqref{eq:definition_MMSE}, the claim follows.
\end{IEEEproof}
We are now ready to present our new lower bound. 
\begin{theorem}\label{thm:Theorem_bound}
    Assume the three following conditions hold:
    \begin{enumerate}
        \item For all $\bfx \in \cX$ the distribution $P_{\sfY | \sfX = \bfx}$ has the pdf of the form in~(\ref{eq:exp_family}) and it satisfies a Poincaré inequality with respect to $(\cA , \kappa(\bfx))$. \label{bound_cond_1}
        \item $\bfy \mapsto i_{P_{\sfX, \sfY}}(\bfx, \bfy) \in \cA$ for every $\bfx$ such that $\kappa(\bfx) > 0$. \label{bound_cond_2}
        \item There exists a $\rho(\bfx) \geq 0$ such that  $\sigma_{\text{min}}((\bfJ_{\sfy} \bfT(\sfy))^{+}) \geq \rho(\bfx)$ for all $\bfy \in \cY$. \label{bound_cond_3}
    \end{enumerate}
    Then,
    \begin{align}
        \text{mmse}(\eta(\sfX)| \sfY) \geq \bbE \left[\rho^{2}(\bfx) \kappa(\sfX) \variance{i_{P_{\sfX, \sfY}}(\sfX, \sfY) | \sfX}\right]. \label{eq:Poi_bound}
    \end{align}
\end{theorem}
\begin{IEEEproof}
    We have
    \begin{align}
        &\text{mmse}(\eta(\sfX) |\sfY) = \bbE [\| \eta(\sfX)- \bbE[\eta(\sfX) | \sfY]\|^{2}] \label{eq:mmse}\\
        &= \bbE \left[\| (\bfJ_{\sfY} \bfT(\sfY))^{+} \nabla_{\sfY} i_{P_{\sfX, \sfY}} (\sfX , \sfY) \|^{2}\right] \label{eq:use_Theorem_mmse_Jacobian}\\
        &\geq \bbE \left[\rho^{2}(\sfX) \| \nabla_{\sfY} i_{P_{\sfX, \sfY}} (\sfX, \sfY) \|^{2}\right] \label{eq:use_Theorem_bound} \\
        &= \expect{ \rho^{2}(\sfX) \expcnd{\| \nabla_{\sfY} i_{P_{\sfX, \sfY}} (\sfX, \sfY) \|^{2}}{\sfX} }  \\
        &\geq \bbE \left[ \rho^{2}(\sfX)\kappa(\sfX) \variance{i_{P_{\sfX, \sfY}}(\sfX , \sfY)| \sfX}\right] \label{eq:use_Poincare},
    \end{align}
    where~\eqref{eq:use_Theorem_mmse_Jacobian} follows from applying Theorem~\ref{Theorem_mmse_Jacobian};~\eqref{eq:use_Theorem_bound} from employing condition~\ref{bound_cond_3} and the inequality $\| \bfA \bfx \| \geq \sigma_{\text{min}}(\bfA) \|\bfx \|$; and~\eqref{eq:use_Poincare} from using the Poincaré inequality~\eqref{eq:Poincare_ineq}  and conditions~\ref{bound_cond_1} and~\ref{bound_cond_2}. This concludes the proof.
\end{IEEEproof}
\section{Channels with Blockage Probability}\label{Section:channel_with_blockage}
In this section, we use our new bound in Theorem~\ref{thm:Theorem_bound} 
for the problem of channel estimation in ISAC systems. We focus on Gaussian channels with blockage probability, and we derive MMSE lower bounds for both scalar and vector cases.

\subsection{System Model}
Let us consider $T$ uses of an $M\times N$ multiple-input multiple-output (MIMO) channel.  The model for the sensing subsystem of an ISAC system is
\begin{align}\label{eq:sys_model}
    \bfY = \bfH \bfX + \bfZ,
\end{align}
where $\bfY \in \bbC^{N \times T}$ is the sensing received signal, $\bfX \in \bbC^{M \times T}$ is the transmitted signal, $\bfH \in \bbC^{N \times M}$ is the target channel response matrix, and $\bfZ \in \bbC^{N \times T}$ is the sensing noise modeled as independent and identically distributed (i.i.d.) circularly symmetric complex Gaussian random variable with zero mean, namely $\text{vec}(\bfZ) \sim {\cal CN}({\bf0}, \sigma_{\rm s}^2 {\sfI}_{NT})$. 
The sample covariance matrix of the transmitted signal, $\bfR_{\bfX} \triangleq T^{-1}\bfX \bfX^{\dagger}$, will play a central role in both the analysis of the sensing system's performance and in imposing an average-power constraint $P_T$ for the communication problem, i.e., $\tr{ \bfR_{\bfX} } \leq M P_T$. We assume $\bfX$ to be known at the sensing receiver, which observes $\bfY$. The sensing performance is the MSE
\begin{align}
    \text{MSE} &\triangleq \expect{\|\bfH - \hat{\bfH} \|^2},
    \label{eq:MSE of A def}
\end{align}
where $\hat{\bfH}$ is an estimate of $\bfH$ based on $(\bfY, \bfX)$.
The optimal MSE estimator is 
$\expcnd{\bfH}{\bfY, \bfX}$.   

In the rest of this work, we consider the following complex to real mapping for all column vectors and matrices 
\begin{align} \label{eq:ReIm_matrix}
\overline{\bfx} = 
    \begin{bmatrix}
        \text{Re}({\bfx}) \\
        \text{Im}({\bfx})
    \end{bmatrix},\,
    \overline{\bfX} = 
    \begin{bmatrix}
        \text{Re}({\bfX}) & -\text{Im}({\bfX}) \\
        \text{Im}({\bfX}) & \text{Re}({\bfX})
    \end{bmatrix}.
\end{align}
Note that in this mapping $\bfR_{\overline{\bfX^\sfT}} \triangleq (2T)^{-1}(\overline{\bfX^\sfT})^{\dagger} \overline{\bfX^\sfT}$.
It can be shown that~\eqref{eq:sys_model} can be rewritten in vector form as
\begin{align}
    \overline{\vect{\bfY^\sfT}} = (I_{N} \otimes \overline{\bfX^\sfT})  \overline{\vect{\bfH^\sfT}} +  \overline{\vect{\bfZ^\sfT}}. \label{eq:reim_vect_sensing}
\end{align}
Let us introduce the quantities $\sfY \triangleq \overline{\vect{\bfY^\sfT}}$, $\sfH \triangleq \overline{\vect{\bfH^\sfT}}$, $\sfZ \triangleq \overline{\vect{\bfZ^\sfT}}$, and $\sfC_{\bfX} \triangleq I_{N} \otimes \overline{\bfX^\sfT}$ so that the complex vectorized sensing model \eqref{eq:reim_vect_sensing} becomes
\begin{align} \label{eq:new_notation_model}
    \sfY = \sfC_{\bfX} \sfH + \sfZ.
\end{align}
In the next subsection, we start with the derivation of our new Poincaré lower bound on the MMSE for the (real) scalar case, i.e., the real part of~\eqref{eq:sys_model} when $N=M=T=1$.
\subsection{Real-Valued Scalar Model}
Consider having channel fading with pdf
\begin{align}
f_\sfH(x) = (1-\alpha)\delta(x) + \alpha \mathcal{N}(x; 0,\sigma_\sfH^2),
\end{align}
where $\alpha \in (0,1]$ and $1-\alpha$ is the probability of having a blockage. In the following, we derive the Poincaré lower bound on the MMSE of $\sfH$. Then, we demonstrate that in the high SNR regime the bound is a function of $\sfX^2$.
Finally, we present the MMSE computed using Theorem~\ref{Theorem_mmse_Jacobian} and the LMMSE. 

In order to 
use Theorem~\ref{thm:Theorem_bound}, we need to evaluate the information density. Starting from
\begin{align}
    &f_{\sfY|\sfH,\sfX}(y|h,x) = \frac{1}{\sqrt{2\pi \sigma_{\rm s}^2}} e^{ -\frac{1}{2} \frac{(y-hx)^2}{\sigma_{\rm s}^2} },\\
    &f_{\sfY|\sfX}(y|x) = (1-\alpha)\frac{e^{ -\frac{1}{2} \frac{y^2}{\sigma_{\rm s}^2} }}{\sqrt{2\pi \sigma_{\rm s}^2}} + \alpha \frac{e^{ -\frac{1}{2} \frac{y^2}{(x^2\sigma_\sfH^2 + \sigma_{\rm s}^2)} }}{\sqrt{2\pi (x^2\sigma_\sfH^2 + \sigma_{\rm s}^2)}},
\end{align}
we get
\begin{align}
    &i_{P_{\sfH,\sfX,\sfY}} (h;y,x) \nonumber\\
    &= \log \left(\frac{f_{\sfY|\sfH,\sfX}(y|h,x)}{f_{\sfY|\sfX}(y|x)} \right) \\
    &= \frac{-(y-hx)^2+(y)^2}{2\sigma_{\rm s}^2} \\
    &\quad-\log \left(1-\alpha + \alpha \sqrt{\frac{\sigma_{\rm s}^2}{x^2\sigma_\sfH^2 + \sigma_{\rm s}^2}} e^{ \frac{y^2}{2\sigma_{\rm s}^2} \frac{x^2\sigma_\sfH^2}{x^2 \sigma_\sfH^2+\sigma_{\rm s}^2} } \right).
\end{align}
We have $\bfT(y) = \frac{yx}{\sigma_{\rm s}^2}$, and by using condition~\ref{bound_cond_3} in Theorem~\ref{thm:Theorem_bound} we get $\rho = \frac{\sigma_{\rm s}^2}{x}$. Since $h(y) = \frac{e^{-\frac{y^2}{2\sigma_{\rm s}^2}}}{\sqrt{2\pi \sigma_{\rm s}^2}}$, we have $\kappa(\sfH) = \frac{1}{\sigma_{\rm s}^2}$ by Proposition~\ref{prop:BE_constant}. Combining all the results, the Poincaré lower bound becomes
\begin{align}
    &\text{mmse}(\sfH|\sfY) \geq \bbE[\rho^2 \kappa(\sfH) \varcnd{i_{P_{\sfH,\sfX,\sfY}}(\sfH;\sfY,\sfX)}{\sfH,\sfX}] \\
    &= \bbE \left[\frac{\sigma_{\rm s}^2}{\sfX^2}\mathsf{Var}\left[\frac{\sfY\sfH\sfX}{\sigma_{\rm s}^2}  \right. \right.\nonumber\\
    &\left. \left. \left. - \log \left(1-\alpha + \alpha \sqrt{\frac{\sigma_{\rm s}^2}{\sfX^2\sigma_\sfH^2 + \sigma_{\rm s}^2}} e^{\frac{\sfY^2}{2\sigma_{\rm s}^2}\frac{\sfX^2\sigma_\sfH^2}{\sfX^2 \sigma_\sfH^2+\sigma_{\rm s}^2} } \right) \right| \sfH,\sfX\right] \right] .
    \label{eq:LB_p_on_off}
\end{align}
Now let us evaluate the information density at $y = hx+\sigma_{\rm s}\tilde{\sfZ}$ where $\tilde{\sfZ} \sim {\cal N}(0,1)$; we get
\begin{align}
    & \log\sqrt{2\pi \sigma^2_s}+ i_{P_{\sfH,\sfX,\sfY}} (h; hx+\sigma_{\rm s}\tilde{\sfZ},x) \nonumber\\
    &= -\frac{1}{2}  \tilde{\sfZ}^2 -\log \left((1-\alpha)\frac{e^{ -\frac{1}{2} \left( \frac{hx}{\sigma_{\rm s}}+\tilde{\sfZ}\right)^2 }}{\sqrt{2\pi \sigma_{\rm s}^2}} + \alpha \frac{e^{ -\frac{1}{2} \frac{( hx+\sigma_{\rm s}\tilde{\sfZ})^2}{x^2\sigma_\sfH^2 + \sigma_{\rm s}^2} }}{\sqrt{2\pi (x^2\sigma_\sfH^2 + \sigma_{\rm s}^2)}} \right). \label{eq:shifted_info_den}
\end{align}
For $x \ne 0$ and $h\ne 0$, 
the high SNR behavior (i.e., as $\sigma_{\rm s} \to 0$) of the information density  is
\begin{align}
     &\lim_{\sigma_{\rm s} \to 0}\log\sqrt{2\pi \sigma^2_s}+ i_{P_{\sfH,\sfX,\sfY}} (h; hx+\sigma_{\rm s}\tilde{\sfZ},x) \nonumber\\
     &= -\frac{1}{2}  \tilde{\sfZ}^2 -\log \left( \alpha \frac{1}{\sqrt{2\pi (x^2\sigma_\sfH^2 )}} \exp \left( -\frac{1}{2} \frac{ h^2}{\sigma_\sfH^2 } \right) \right). 
     \label{eq:limit_small_sigma_s_xh_not0}
\end{align}
If $x \ne 0$ and $h = 0$ then
\begin{align}
     \lim_{\sigma_{\rm s} \to 0} i_{P_{\sfH,\sfX,\sfY}} (h; hx+\sigma_{\rm s}\tilde{\sfZ},x) &= -\log(1-\alpha),
\end{align}
while for $x = 0$ we have
\begin{align}
     \lim_{\sigma_{\rm s} \to 0} i_{P_{\sfH,\sfX,\sfY}} (h; hx+\sigma_{\rm s}\tilde{\sfZ},x) &= 0.
\end{align}
To correctly evaluate the high SNR behavior of the variance of the information density, we need the following result.
\begin{lemma}\label{lem:dct}
     For $x\ne 0$ and $h \ne 0$, we have
     \begin{align}
         &\lim_{\sigma_{\rm s} \to 0} \variance{\log\sqrt{2\pi \sigma^2_s}+i_{P_{\sfH,\sfX,\sfY}} (h; hx+\sigma_{\rm s}\tilde{\sfZ},x)} \nonumber \\
         &= \variance{\lim_{\sigma_{\rm s} \to 0}\log\sqrt{2\pi \sigma^2_s}+i_{P_{\sfH,\sfX,\sfY}} (h; hx+\sigma_{\rm s}\tilde{\sfZ},x)}.
     \end{align}
 \end{lemma}
 \begin{IEEEproof}
     First of all, by using \eqref{eq:shifted_info_den} and triangle inequality we have
     \begin{align}
        & \left| \log\sqrt{2\pi \sigma^2_s}+i_{P_{\sfH,\sfX,\sfY}} (h; hx+\sigma_{\rm s}\tilde{\sfZ},x)\right| \\
        &\le \frac{1}{2}  \tilde{\sfZ}^2 \nonumber\\
        &+\left|\log \left((1-\alpha)\frac{e^{ -\frac{1}{2}  \frac{( hx+\sigma_{\rm s}\tilde{\sfZ})^2}{ \sigma_{\rm s}^2} }}{\sqrt{2\pi \sigma_{\rm s}^2}} + \alpha \frac{e^{ -\frac{1}{2} \frac{( hx+\sigma_{\rm s}\tilde{\sfZ})^2}{x^2\sigma_\sfH^2 + \sigma_{\rm s}^2} }}{\sqrt{2\pi (x^2\sigma_\sfH^2 + \sigma_{\rm s}^2)}} \right) \right|.
     \end{align}
Next, notice that
     \begin{align}
         &\lim_{\sigma_{\rm s} \to 0} \log \left((1-\alpha)\frac{e^{ -\frac{1}{2}  \frac{( hx+\sigma_{\rm s}\tilde{\sfZ})^2}{ \sigma_{\rm s}^2} }}{\sqrt{2\pi \sigma_{\rm s}^2}} + \alpha \frac{e^{ -\frac{1}{2} \frac{( hx+\sigma_{\rm s}\tilde{\sfZ})^2}{x^2\sigma_\sfH^2 + \sigma_{\rm s}^2} }}{\sqrt{2\pi (x^2\sigma_\sfH^2 + \sigma_{\rm s}^2)}}  \right) \\
         &= \log \left( \alpha \frac{1}{\sqrt{2\pi x^2\sigma_\sfH^2 }} \exp \left( -\frac{1}{2} \frac{ h^2}{\sigma_\sfH^2 } \right) \right)< \infty
     \end{align}
     a.s. in $\tilde{\sfZ}$. This means that for all sufficiently small $\sigma_{\rm s}$, we have that 
     \begin{align}
        & \left| \log\sqrt{2\pi \sigma^2_s}+i_{P_{\sfH,\sfX,\sfY}} (h; hx+\sigma_{\rm s}\tilde{\sfZ},x)\right| \le \frac{1}{2}  \tilde{\sfZ}^2 + c,
     \end{align}
     a.s. in $\tilde{\sfZ}$, where $c$ is a finite constant independent of $\sigma_{\rm s}$. Then, by using the dominated convergence theorem, we finally prove the claim.
 \end{IEEEproof}
For $x \ne 0$ and $h\ne 0$, the variance of the information density has the following asymptotic behavior 
\begin{align}
    &\lim_{\sigma_{\rm s} \to 0} \variance{i_{P_{\sfH,\sfX,\sfY}} (h; hx+\sigma_{\rm s}\tilde{\sfZ},x)} \nonumber\\
    &= \lim_{\sigma_{\rm s} \to 0} \variance{\log\sqrt{2\pi \sigma^2_s}+i_{P_{\sfH,\sfX,\sfY}} (h; hx+\sigma_{\rm s}\tilde{\sfZ},x)}  \\
    &= \variance{\lim_{\sigma_{\rm s} \to 0}\log\sqrt{2\pi \sigma^2_s}+i_{P_{\sfH,\sfX,\sfY}} (h; hx+\sigma_{\rm s}\tilde{\sfZ},x)} \\
    &=\frac{1}{2} \nonumber 
\end{align}
where the last step follows from Lemma \ref{lem:dct}. 
In the other cases, namely when $\{x=0\} \vee \{x \ne 0, \, h=0\}$, we have
\begin{align}
     \lim_{\sigma_{\rm s} \to 0} \variance{i_{P_{\sfH,\sfX,\sfY}} (h; hx+\sigma_{\rm s}\tilde{\sfZ},x)} &=0.
\end{align}
To sum up, by assuming $\sfX \ne 0$, the high SNR asymptotic of the Poincaré lower bound is
\begin{align}
   \lim_{\sigma_{\rm s} \to 0} \frac{\text{mmse}(\sfH|\sfY)}{\sigma^2_{\rm s}} &\geq  
 \lim_{\sigma_{\rm s} \to 0}\bbE\left[\frac{1}{\sfX^2}\varcnd{i_{P_{\sfH,\sfX,\sfY}}(\sfH;\sfY,\sfX)}{\sfH,\sfX}\right] \\
 &=  \Pr(\sfH \ne 0) \frac{1}{2} \expect{\frac{1}{\sfX^2}}   =   \frac{\alpha}{2} \expect{\frac{1}{\sfX^2}}
 \label{eq:asymptote high SNR scalar}
\end{align}
which shows a dependence on the sample covariance $\sfX^2$. 
 
\begin{figure}
    \centering
    \includegraphics[width=0.5\textwidth]{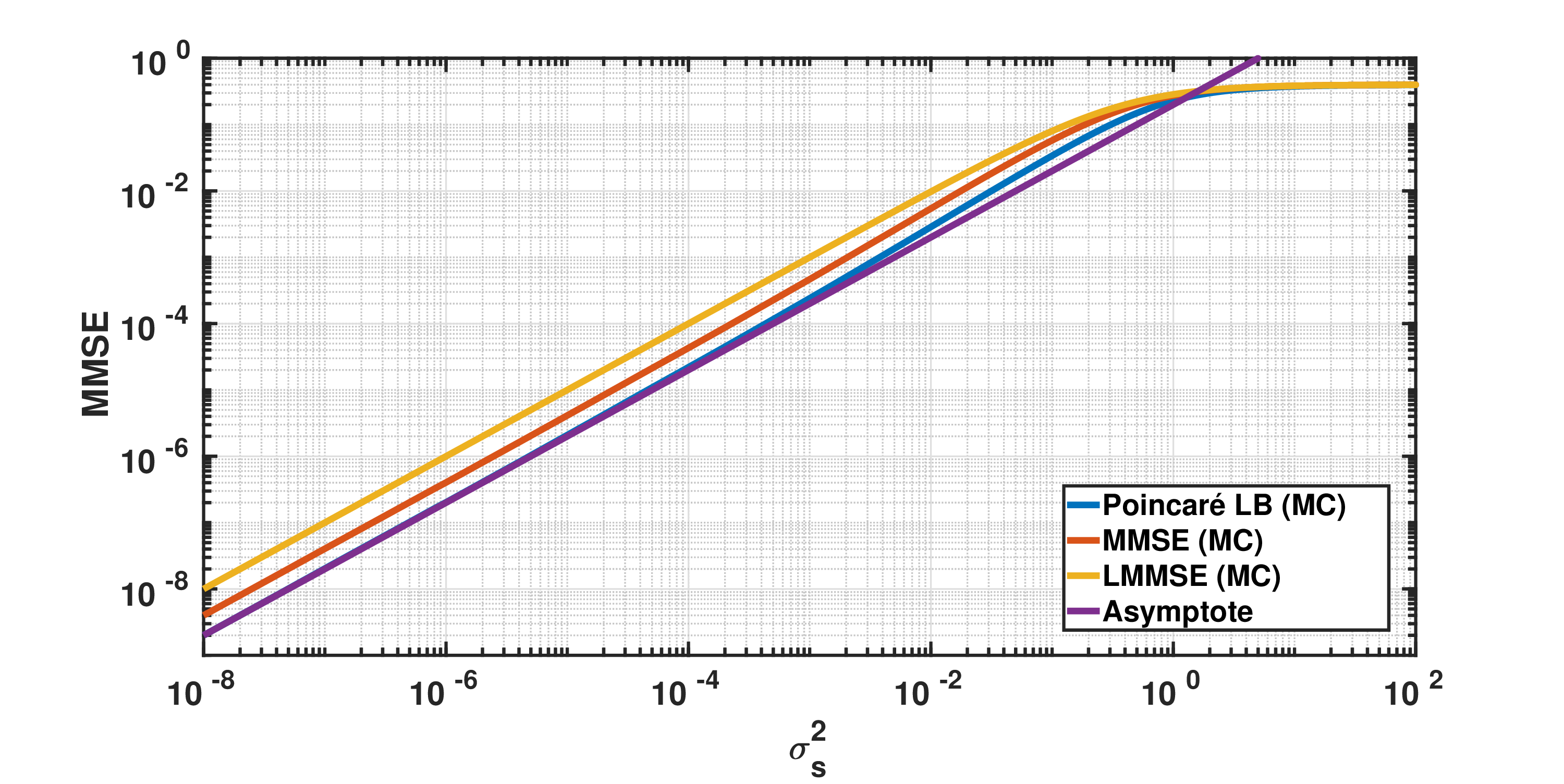}
    \caption{Poincaré lower bound in~\eqref{eq:LB_p_on_off}, MMSE in~\eqref{eq:MMSE_scalar}, and LMMSE in~\eqref{eq:LMMSE_scalar} for different values of $\sigma_{\rm s}^2$ with $\sfX=1$, $\alpha=0.4$, and $ \sigma_h^2=1$. For each value of $\sigma_{\rm s}^2$, the simulations are run for $10^8$ Monte Carlo trials.} 
    \label{fig:LP_p_alpha_0.4}
\end{figure}

Notice that, for this channel model the MMSE can be evaluated by Theorem~\ref{Theorem_mmse_Jacobian} and is given by
\begin{align}\label{eq:MMSE_scalar}
    &\text{mmse}(\sfH|\sfY) = \bbE\left[\frac{\sigma_{\rm s}^4}{\sfX^2}\bbE\left[\|\nabla_{\sfY} i_{P_{\sfH,\sfY,\sfX}}(\sfH; \sfY,\sfX)\|^2|\sfH,\sfX\right]\right],
\end{align}
where the gradient of the information density is
\begin{align}
    &\nabla_{y} i_{P_{\sfH,\sfY,\sfX}}(h; y,x) =
     \frac{hx-y}{\sigma_{\rm s}^2} \nonumber\\
     &+ \frac{\frac{(1-\alpha)y}{\sigma_{\rm s}^2\sqrt{2\pi\sigma_{\rm s}^2}}e^{-\frac{y^2}{2\sigma_{\rm s}^2}} + \frac{\alpha y}{(x^2\sigma_\sfH^2 + \sigma_{\rm s}^2)\sqrt{2\pi (x^2\sigma_\sfH^2 + \sigma_{\rm s}^2)}} e^{ -\frac{1}{2} \frac{y^2}{x^2\sigma_\sfH^2 + \sigma_{\rm s}^2} }}{{\frac{(1-\alpha)}{\sqrt{2\pi \sigma_{\rm s}^2}} e^{ -\frac{1}{2} \frac{y^2}{\sigma_{\rm s}^2} } + \frac{\alpha}{\sqrt{2\pi (x^2\sigma_\sfH^2 + \sigma_{\rm s}^2)}} e^{ -\frac{1}{2} \frac{y^2}{x^2\sigma_\sfH^2 + \sigma_{\rm s}^2} }}}.
\end{align}
The LMMSE is given by
\begin{align}\label{eq:LMMSE_scalar}
    \text{lmmse}(\sfH|\sfY) = \bbE\left[ \frac{\alpha \sigma_{H}^2\sigma_{\rm s}^2}{(\alpha \sfX^2\sigma_{\sfH}^2+\sigma_{\rm s}^2)} \right].
\end{align}
In \figurename~\ref{fig:LP_p_alpha_0.4}, we plot the Poincaré lower bound in~\eqref{eq:LB_p_on_off}, the MMSE in~\eqref{eq:MMSE_scalar}, and the LMMSE in~\eqref{eq:LMMSE_scalar} for different values of $\sigma_{\rm s}^2$ by using Monte Carlo simulation. The figure shows the tightness of the bound and confirms the asymptotic behavior of the curve as ${\sigma_{\rm s} \to 0}$, i.e., in the high SNR regime. 

\subsection{Vector Model}
The distribution of the channel $\sfH$ in the vector case is 
\begin{align}
f_\sfH(\bfx) = (1-\alpha) \delta(\bfx) + \alpha \mathcal{N}\left(\bfx;{\bf0},\frac{\sigma_{\sfH}^2}{2} I_{2NM}\right).
\end{align}
Here, the $N$ receive sensing antennas experience a blockage event that lasts $T$ channel uses with probability $1-\alpha$.
As for the scalar model, we compute the Poincaré lower bound and then present the MMSE and the LMMSE. 
First, we compute the information density for the vector case. We have
\begin{align}
    {\mu}_{\sfY|\sfH,\bfX} &\triangleq \expcnd{\sfY}{\sfH,\bfX}  = \sfC_{\bfX}\sfH \\
    {\mu}_{\sfY|\bfX} &\triangleq \expcnd{\sfY}{\bfX} = \sfC_{\bfX}\expect{\sfH} = \mathbf{0}
\end{align}
and
\begin{align}
    {\bf\Sigma}_{\sfY |\sfH, \bfX} &\triangleq \frac{\sigma_{\rm s}^2}{2} I_{2NT} \\
    {\bf\Sigma}_{\sfY |\bfX} &\triangleq \sfC_{\bfX}\sfC_{\bfX}^\sfT \frac{\sigma^2_\sfH}{2}+\frac{\sigma_{\rm s}^2}{2} I_{2NT}.
\end{align}
Therefore, we have
\begin{align}
    &f_{\sfY | \sfH,\bfX}(\sfy |\sfH, \bfX) = \frac{e^{-\frac{1}{2}(\sfy-{\mu}_{\sfY|\sfH,\bfX})^{\sfT} {\bf\Sigma}_{\sfY | \sfH,\bfX}^{-1} (\sfy-{\mu}_{\sfY|\sfH,\bfX})}}{\det(2\pi {\bf\Sigma}_{\sfY |\sfH, \bfX})^{1/2}}, 
\end{align}
and
\begin{align}
    f_{\sfY | \bfX}(\sfy | \bfX) &= (1-\alpha)\frac{ e^{-\frac{\sfy^{\sfT} \sfy}{\sigma_{\rm s}^2}}}{\det(\pi \sigma_{\rm s}^2 I_{2NT})^{1/2}} \nonumber\\
    &\quad + \alpha \frac{ e^{-\frac{1}{2}(\sfy-{\mu}_{\sfY|\bfX})^{\sfT} {\bf\Sigma}_{\sfY | \bfX}^{-1} (\sfy-{\mu}_{\sfY|\bfX})}}{\det(2\pi {\bf\Sigma}_{\sfY | \bfX})^{1/2}}.
\end{align}
If then follows that the information density is
\begin{align}
    &i_{P_{\sfH,\bfX,\sfY}}(\sfH; \sfy, \bfX) = \log\left(\frac{f_{\sfY | \sfH,\bfX}(\sfy | \sfH,\bfX)}{f_{\sfY | \bfX}(\sfy | \bfX)}\right) \\
    &= \log\left( \frac{e^{-\frac{1}{2}(\sfy-{\mu}_{\sfY|\sfH,\bfX})^{\sfT} {\bf\Sigma}_{\sfY | \sfH,\bfX}^{-1} (\sfy-{\mu}_{\sfY|\sfH,\bfX})}}{\det(2\pi {\bf\Sigma}_{\sfY |\sfH, \bfX})^{1/2}}\right) \nonumber \\
    &- \log \left( \frac{(1-\alpha) e^{-\frac{\sfy^{\sfT} \sfy}{\sigma_{\rm s}^2}}}{\det(\pi \sigma_{\rm s}^2 I_{2NT})^{1/2}} + \frac{\alpha e^{-\frac{1}{2}(\sfy-{\mu}_{\sfY|\bfX})^{\sfT} {\bf\Sigma}_{\sfY | \bfX}^{-1} (\sfy-{\mu}_{\sfY|\bfX})}}{\det(2\pi {\bf\Sigma}_{\sfY | \bfX})^{1/2}} \right).\label{eq:info_density_vect}
\end{align}
By applying Proposition~\ref{prop:BE_constant}, we find $\kappa(\sfH) = \frac{2}{\sigma_{\rm s}^2} I_{2NT}$. 
Next, we calculate the value of $\rho$ as
\begin{align}
    \rho=\sigma_{\min}((\bfJ_{\bfy} \bfT(\bfy))^{+}) &= \sigma_{\min}\left(\frac{\sigma_{\rm s}^2}{2}\sfC_{\bfX}^{+}\right). \label{eq:rho_blockage}
\end{align}
By substituting~\eqref{eq:info_density_vect}, $\kappa(\sfH)$, and $\rho$ in~\eqref{eq:Poi_bound} we obtain the Poincaré lower bound.

The information density at $\sfy = \sfC_{\bfX} \sfH + \sigma_{s} \tilde{\sfZ}$ where $\tilde{\sfZ} \sim {\cal N}({\bf0},I_{2NT})$ is
\begin{align}
    &\log(\det(2\pi {\bf\Sigma}_{\sfY |\sfH, \bfX})^{1/2}) + i_{P_{\sfH,\bfX,\sfY}}(\sfH; \sfC_{\bfX} \sfH + \sigma_{s} \tilde{\sfZ}, \bfX) \nonumber\\
    &= - \tilde{\sfZ}^{\sfT} \tilde{\sfZ} - \log \left( \frac{(1-\alpha) e^{-\left(\frac{\sfC_{\bfX} \sfH}{\sigma_{\rm s}} + \tilde{\sfZ}\right)^{\sfT} \left(\frac{\sfC_{\bfX} \sfH}{\sigma_{\rm s}} + \tilde{\sfZ}\right)}}{\det(\pi \sigma_{\rm s}^2 I_{2NT})^{1/2}}\right. \nonumber\\
    &\left.\quad + \frac{\alpha e^{-\frac{1}{2}(\sfC_{\bfX} \sfH + \sigma_{\rm s} \tilde{\sfZ})^{\sfT} \left(\sfC_{\bfX}\sfC_{\bfX}^\sfT \frac{\sigma^2_\sfH}{2}+\frac{\sigma_{\rm s}^2}{2} I_{2NT}\right)^{-1} (\sfC_{\bfX} \sfH + \sigma_{\rm s} \tilde{\sfZ})}}{\det\left(2\pi \left(\sfC_{\bfX}\sfC_{\bfX}^\sfT \frac{\sigma^2_\sfH}{2}+\frac{\sigma_{\rm s}^2}{2} I_{2NT}\right)\right)^{1/2}} \right).\label{eq:shifted_info_den_vect}
\end{align}
 When $\sfC_{\bfX} \ne {\bf0}$ and $\sfH \ne {\bf0}$, the high SNR behavior of the information density as $\sigma_{\rm s} \to 0$ is as follows
\begin{align}
    &\lim_{\sigma_{\rm s} \to 0}\log(\det(2\pi {\bf\Sigma}_{\sfY |\sfH, \bfX})^{1/2}) + i_{P_{\sfH,\bfX,\sfY}}(\sfH; \sfC_{\bfX} \sfH + \sigma_{s} \tilde{\sfZ}, \bfX)  \nonumber\\
    &=- \tilde{\sfZ}^{\sfT} \tilde{\sfZ} -\log \left( \alpha \det\left(\pi\sfC_{\bfX}\sfC_{\bfX}^\sfT \sigma^2_\sfH\right)^{-1/2} \exp{\left(- \frac{\sfH^{\sfT} \sfH}{\sigma_{\sfH}^2}\right)} \right).
    \label{eq:limit_small_sigma_s_xh_not0_vect}
\end{align}
If $\sfC_{\bfX} \ne {\bf0}$ and $\sfH = {\bf0}$, then
\begin{align}
    \lim_{\sigma_{\rm s} \to 0} i_{P_{\sfH,\bfX,\sfY}}(\sfH; \sfC_{\bfX} \sfH + \sigma_{s} \tilde{\sfZ}, \bfX) = - \log(1 - \alpha), 
\end{align}
while for $\sfC_{\bfX} = {\bf0}$ we have
\begin{align}
    \lim_{\sigma_{\rm s} \to 0} i_{P_{\sfH,\bfX,\sfY}}(\sfH; \sfC_{\bfX} \sfH + \sigma_{s} \tilde{\sfZ}, \bfX) = 0.
\end{align}
For $\sfC_{\bfX} \ne {\bf0}$ and $\sfH \ne {\bf0}$, the variance of the information density has the following asymptotic behavior
\begin{align}
    &\lim_{\sigma_{\rm s} \to 0} \variance{i_{P_{\sfH,\bfX,\sfY}}(\sfH; \sfC_{\bfX} \sfH + \sigma_{s} \tilde{\sfZ}, \bfX)} \nonumber\\
    &= \lim_{\sigma_{\rm s} \to 0} \variance{\log(\det(2\pi {\bf\Sigma}_{\sfY |\sfH, \bfX})^{1/2}) + i_{P_{\sfH,\bfX,\sfY}}(\sfH;\sfY, \bfX)} \nonumber\\
    &= \variance{\lim_{\sigma_{\rm s} \to 0}\log(\det(2\pi {\bf\Sigma}_{\sfY |\sfH, \bfX})^{1/2}) + i_{P_{\sfH,\bfX,\sfY}}(\sfH; \sfY, \bfX)} \label{eq:apply_DCT_vect}\\
    &= \frac{2}{\sigma_{\rm s}^4} \text{tr} \left[\left(\frac{\sigma_{\rm s}^2}{2}I_{2NT}\right)^2\right] = NT,
\end{align}
where \eqref{eq:apply_DCT_vect} follows from a straightforward extension of Lemma~\ref{lem:dct} to vector space. 
In the other cases, namely $\{\sfC_{\bfX}={\bf0}\} \vee \{\sfC_{\bfX} \ne {\bf0}, \, \sfH={\bf0}\}$, we have
\begin{align}
    \lim_{\sigma_{\rm s} \to 0} \variance{i_{P_{\sfH,\bfX,\sfY}}(\sfH; \sfC_{\bfX} \sfH + \sigma_{\rm s}\tilde{\sfZ}, \bfX)} = 0. 
\end{align}
Finally, by assuming $\sfC_{\bfX} \ne {\bf0}$, the Poincaré lower bound in the vector case is asymptotically equal to
\begin{align}
    &\lim_{\sigma_{\rm s} \to 0} \frac{\text{mmse}(\sfH|\sfY)}{\sigma^2_s} \nonumber\\
    &\geq \lim_{\sigma_{\rm s} \to 0} \frac{1}{\sigma^2_s} \bbE[\rho^2 \kappa(\sfH) \varcnd{i_{P_{\sfH,\sfX,\sfY}}(\sfH;\sfY,\sfX)}{\sfH,\sfX}] \\
    &\geq \lim_{\sigma_{\rm s} \to 0} \frac{1}{\sigma^2_s} \bbE\left[\left\|\frac{2}{\sigma_{\rm s}^2}\sfC_{\bfX}\right\|^{-2} \frac{2}{\sigma_{\rm s}^2} \varcnd{i_{P_{\sfH,\sfX,\sfY}}(\sfH;\sfY,\sfX)}{\sfH,\sfX}\right] \label{eq:asympotot_Forb_norm}\\
    &= \Pr(\sfH \ne {\bf 0}) \frac{1}{4} \text{tr}^{-1} \left(\bbE\left[{\bfR_{\overline{\bfX^\sfT}}}\right]\right)
    = \frac{\alpha}{4} \text{tr}^{-1} \left(\bbE\left[{\bfR_{\overline{\bfX^\sfT}}}\right]\right),
    \label{eq:asympotot_slop}
\end{align}
where in~\eqref{eq:asympotot_Forb_norm} we used \eqref{eq:rho_blockage} and $\sigma_{\min}(\bfA^{+})= \sigma_{\max}^{-1}(\bfA) \ge (\sum_{i} \sigma_{i}(\bfA))^{-1} = \sqrt{\text{tr}^{-1}(\bfA^{\dagger} \bfA)}$. Similar to the scalar case, \eqref{eq:asympotot_slop} shows that asymptotically the bound is a function of the sample covariance matrix ${\bfR_{\overline{\bfX^\sfT}}}$.

By using Theorem~\ref{Theorem_mmse_Jacobian},  the MMSE is
\begin{align}\label{eq:MMSE_vector_case}
    &\text{mmse}(\sfH|\sfY)= \nonumber\\
    &\bbE\left[\left\|\frac{\frac{(1-\alpha) \sfH e^{-\frac{\sfy^{\sfT} \sfy}{\sigma_{\rm s}^2}}}{\det(\pi \sigma_{\rm s}^2 I_{2NT})^{1/2}} + \frac{\alpha \left(-\frac{\sigma^2_{\sfH}}{2}\sfC_{\bfX}^\sfT{\bf\Sigma}_{\sfY | \bfX}^{-1}\sfy+\sfH\right) e^{-\frac{1}{2}\sfy^{\sfT} {\bf\Sigma}_{\sfY | \bfX}^{-1} \sfy}}{\det(2\pi {\bf\Sigma}_{\sfY | \bfX})^{1/2}} }{ \frac{(1-\alpha) e^{-\frac{\sfy^{\sfT} \sfy}{\sigma_{\rm s}^2}}}{\det(\pi \sigma_{\rm s}^2 I_{2NT})^{1/2}} + \frac{\alpha e^{-\frac{1}{2}\sfy^{\sfT} {\bf\Sigma}_{\sfY | \bfX}^{-1} \sfy}}{\det(2\pi {\bf\Sigma}_{\sfY | \bfX})^{1/2}}}\right\|^2\right],
\end{align}
and the LMMSE is
\begin{align}\label{eq:LMMSE_vector_case}
    \text{lmmse}(\sfH|\sfY,\bfX) &= \frac{N \sigma^2_s}{4T} \bbE\left[ \text{tr}\left(\bfR_{\overline{\bfX^\sfT}}+\frac{\sigma^2_s}{2 \alpha T\sigma^2_\sfH}I_{2M}\right)^{-1} \right].
\end{align}

\begin{figure}
    \centering
    \includegraphics[width=0.5\textwidth]{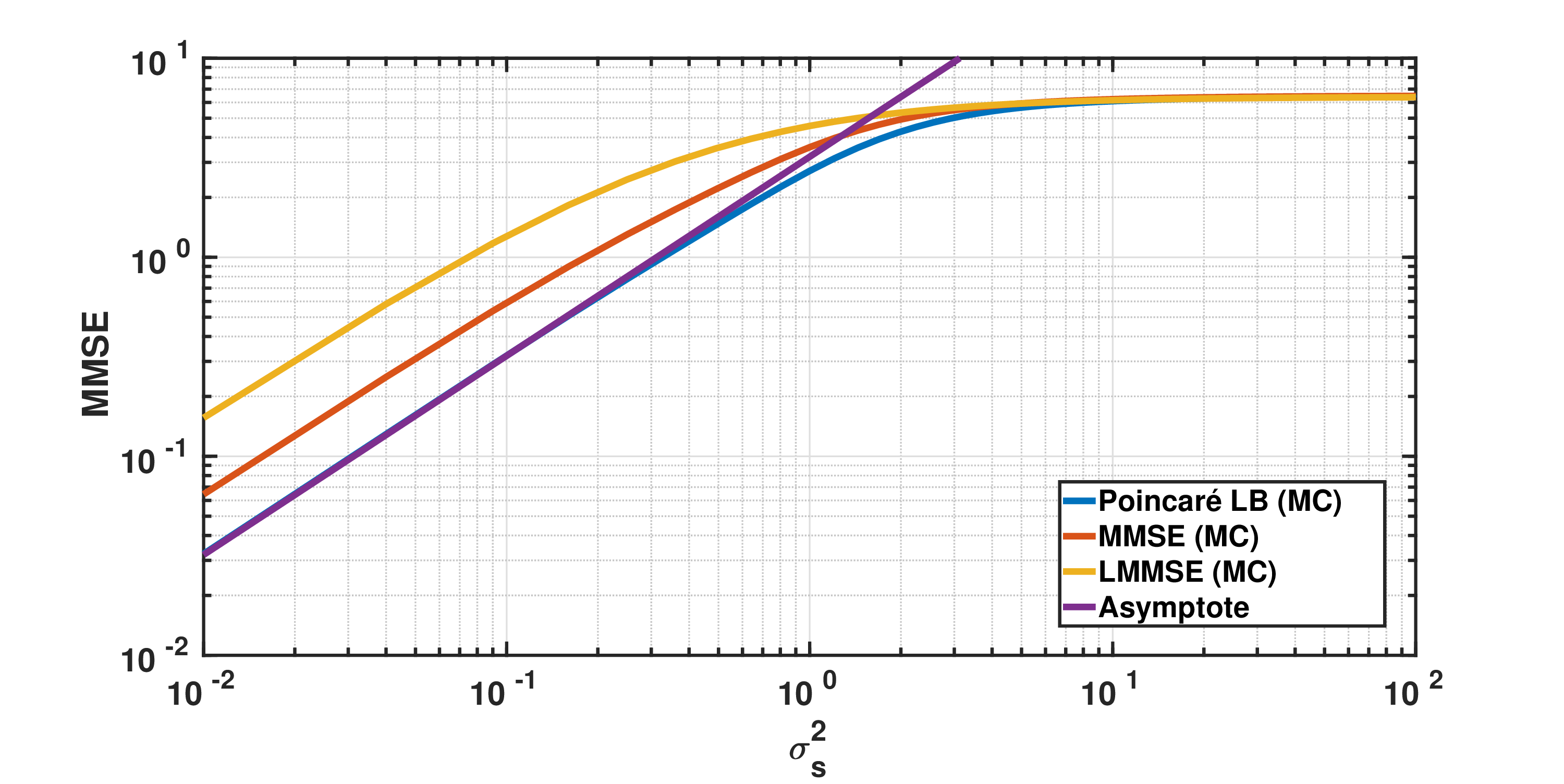}
    \caption{Poincaré lower bound~\eqref{eq:Poi_bound}, MMSE~\eqref{eq:MMSE_vector_case}, and LMMSE~\eqref{eq:LMMSE_vector_case} for $100$ values of $\sigma_{\rm s}^2$, $M = N = T = 4$, $\alpha = 0.4$, $\frac{\sigma_{\sfH}^2}{2} = \frac{1}{2}$, and $\sfC_{\bfX} = I$. For each value of $\sigma_{\rm s}$, the simulations are run for $10^8$ Monte Carlo trials.}
    \label{fig:LB_p_vect_444_alpha_0.4}
\end{figure}

Numerical evaluation results are shown in \figurename~\ref{fig:LB_p_vect_444_alpha_0.4}, from which we observe that the Poincaré lower bound captures the slope of the MMSE at high SNR values. The asymptotic behavior for small values of $\sigma_{\rm s}^2$ is consistent with our theoretical analysis. As we stated before, from~\eqref{eq:asympotot_slop} we can immediately understand the importance of the covariance matrix of the transmitted signal and its effect on the performance of the system. This observation is aligned with what was found in~\cite{ISAC_Caire}, highlighting the practicality of the Poincaré lower bound due to its simpler form compared to the closed form MMSE.

\section{Conclusion}\label{Section:conclusion}
In this work, we proposed a Poincaré lower bound for the non-canonical exponential family of distributions. We calculated the bound for the problem of estimation of channels with blockage probability, which is of great significance for next-generation wireless networks operating at millimeter wave frequencies. In such channels, the Bayesian Cramér-Rao Bound (BCRB) is not applicable due to the fact that the BCRB does not apply to channels with discrete or mixed distributions. As for the channel model, we studied both the real-valued scalar and the complex-valued vector cases. In the high signal-to-noise ratio regime, we demonstrated that the bound is a function of the sample covariance matrix of the transmitted signal. Moreover, we compared the Poincaré bound to the MMSE and the linear MMSE. The main direction for future research will be to utilize the Poincaré lower bound to optimize the transmitted signal in an integrated sensing and communication system for optimal sensing and communication performance. Another promising research path is to obtain a lower bound on the MMSE when the channel matrix is a nonlinear function of a hidden parameter, such as in angle of arrival estimation, when BCRB cannot be calculated.

\bibliographystyle{ieeetr}
\bibliography{bibliography.bib}

\end{document}